\documentclass[conference,a4paper]{APSIPA2021}
\usepackage{multirow}
\usepackage{graphicx}
\usepackage{amsmath}
\usepackage{amsxtra}
\usepackage{threeparttable}

\usepackage{url}
\usepackage{dblfloatfix}
\usepackage{subcaption}
\usepackage{booktabs}
\usepackage{hyperref} 

\usepackage{geometry}
\begin{document}

\title{\emph{Hi,KIA}: A Speech Emotion Recognition Dataset \\ for Wake-Up Words}
\author{%
\authorblockN{%
Taesu Kim\authorrefmark{1}\authorrefmark{4},
SeungHeon Doh\authorrefmark{2}\authorrefmark{4},
Gyunpyo Lee\authorrefmark{1},
Hyungseok Jeon\authorrefmark{3},
Juhan Nam\authorrefmark{2} and
Hyeon-Jeong Suk\authorrefmark{1}
}
\authorblockA{%
\authorrefmark{1}
Department of Industrial Design, KAIST, Daejeon, South Korea \\
E-mail: {tskind77, gyunpyolee, color}@kaist.ac.kr}
\authorblockA{%
\authorrefmark{2}
Graduate School of Culture Technology, KAIST, Daejeon, South Korea\\
E-mail: {seungheondoh, juhan.nam}@kaist.ac.kr}
\authorblockA{%
\authorrefmark{3}
KIA Design Studio, Hyundai Motor Company, Hwaseong, South Korea\\
E-mail: letyoufly@naver.com}
\authorblockA{%
\authorrefmark{4}
Equally contributing authors}
}

\maketitle

\begin{abstract}
Wake-up words (WUW) is a short sentence used to activate a speech recognition system to receive the user's speech input. WUW utterances include not only the lexical information for waking up the system but also non-lexical information such as speaker identity or emotion. In particular, recognizing the user's emotional state may elaborate the voice communication. However, there is few dataset where the emotional state of the WUW utterances is labeled. In this paper, we introduce \emph{Hi, KIA}, a new WUW dataset which consists of 488 Korean accent emotional utterances collected from four male and four female speakers and each of utterances is labeled with four emotional states including anger, happy, sad, or neutral. We present the step-by-step procedure to build the dataset, covering scenario selection, post-processing, and human validation for label agreement. Also, we provide two classification models for WUW speech emotion recognition using the dataset. One is based on traditional hand-craft features and the other is a transfer-learning approach using a pre-trained neural network. These classification models could be used as benchmarks in further research.   
\end{abstract}

\section{Introduction}

\begin{table*}[t]
\centering
\caption{Comparison of some existing public emotion-labeled speech datasets and the proposed Hi,KIA}
\label{tab:relatedData}
\begin{tabular}{lllllll}
\toprule
\textbf{Name} & \textbf{Emotion labels} & \textbf{\# utterances} & \textbf{\# speakers} & \textbf{Accent} & \textbf{Type} & \textbf{Avg.length(s)} \\ \midrule
\begin{tabular}[c]{@{}l@{}}RAVDESS \cite{livingstone2018ryerson} \\ (Audio)\end{tabular} & \begin{tabular}[c]{@{}l@{}}neutral, calm, happy, sad, \\ angry, fearful, disgust, surprised\end{tabular} & 2452 & 24 & \begin{tabular}[c]{@{}l@{}}American\\ (North)\end{tabular} & Sentence & 4.09 \\ \midrule
TESS \cite{tess20} & \begin{tabular}[c]{@{}l@{}}anger, disgust, fear, happiness, \\ pleasant, surprise, sadness, neutral\end{tabular} & 2800 & 2 & \begin{tabular}[c]{@{}l@{}}American\\ (North)\end{tabular} & Sentence & 2.06 \\ \midrule
OK Aura \cite{cambara2022tase} & annoyed, friendly, neutral & \begin{tabular}[c]{@{}l@{}}1247\\ (218) \end{tabular} & 80 & \begin{tabular}[c]{@{}l@{}}Spanish \end{tabular} & Wake-up word & 1.52 \\  \midrule
Hi,KIA (\textbf{Ours}) & \begin{tabular}[c]{@{}l@{}}angry, happy, sad, neutral \end{tabular} & 488 & 8 & Korean & Wake up word & 0.64 \\
\bottomrule
\end{tabular}
\end{table*}

\begin{figure*}[hb]
  \includegraphics[width=\textwidth]{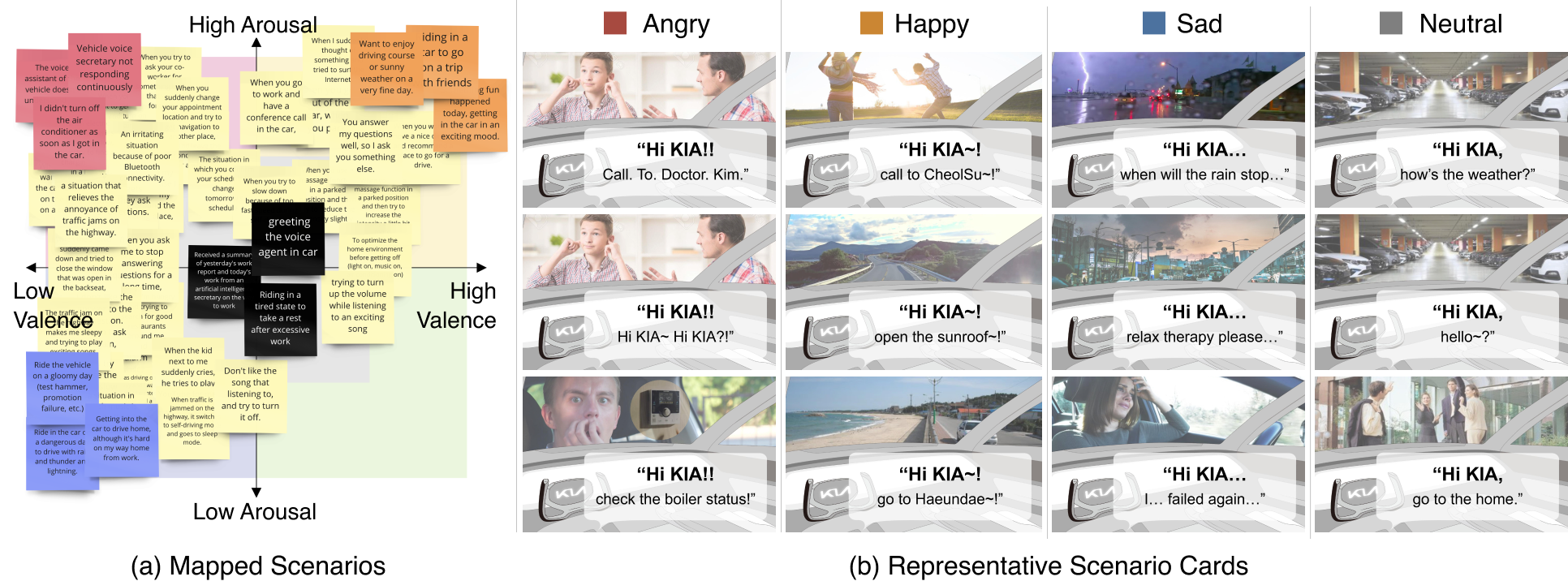}
  \vspace{-4mm}
  \caption{The participants watched with one of the visualized scenarios and pretended to be in the given situation and the emotional circumstance. When ready, they clicked the record button and spoke out the dialogues that began with \emph{Hi, KIA}.}
  \label{fig:corpus}
\end{figure*}

Voice recognition technology has rapidly advanced, deploying voice user interfaces (VUIs) in a wide range of applications. Studies revealed that the VUIs appeal the users because they are invisible \cite{shalini2019development} and the emotional communication is more anticipated than the conventional interface types \cite{nass2005increasing}. In particular, the in-vehicle VUIs have received great interests as they ensure safer driving \cite{strayer2017smartphone, kim2016effects} and enrich the driver's emotional experience \cite{schmidt2020users}.
The most frequent use of VUIs is to speak short wake-up words (WUW) to activate the interface. Previous studies have focused on detecting WUW utterances accurately and quickly \cite{li2019adversarial, gao2020front}. In addition, some of them recognized the speaker's identity to reject imposters from WUW utterances \cite{tsai2021self, qin2020hi}. 
Activating VUIs using WUW is similar to the genuine way of speaking when the user issues a command \cite{vernuccio2020developing}. In many cases, users will speak as if they were talking with another person when giving commands to use VUIs. This naturally conveys the emotion while the user speaks. 
Therefore, understanding the emotions of WUW can provide another dimension to enhance human-machine interaction.



Speech emotion recognition (SER) requires a dataset that includes rich emotional utterances and labels. There are a handful of SER datasets, for example, IEMOCAP \cite{busso2008iemocap}, EmoDB \cite{burkhardt2005database}, RAVDESS \cite{livingstone2018ryerson}, and TESS \cite{tess20}. The SER datasets were designed for text-independent emotion recognition. In other words, the system based on the datasets should recognize the speaker's emotion regardless of the lexical information. On the other hand, the SER dataset for WUW is confined to the signature keyword such as \textit{Ok Google} or \textit{Hey Siri} and, as a result, the length is very short. RAVDESS and TESS have the lexically-matched characteristics. Especially, TESS has the shortest average duration utterance about 2.06 seconds as shown in Table \ref{fig:corpus}. However, WUW utterances are generally even shorter, less or equal to a second. \emph{OK Aura}, a recently released WUW dataset, contains 1247 utterances from 80 speakers with rich metadata annotations such as gender, room size, accent, and emotions \cite{cambara2022tase}. The dataset distinguishes the utterances with three emotions such as annoyed, friendly, or neutral; however, only 218 out of 1247 are labeled.

We constructed a new emotion-labeled WUW dataset, \emph{Hi, KIA}, to address the current limitations. It is composed of 488 recordings from eight Korean voice actors and actresses that correspond to four emotional states including angry, happy, sad and neutral. Compared to the existing datasets, \emph{Hi, KIA} has the shortest average utterance length (0.64 sec) and the emotion of all WUW samples are manually labeled. We present the step-by-step procedure to build the dataset, covering scenario selection, recording/post-processing, and human validation for label agreement. In addition, we provide two classification models for WUW emotion recognition using the dataset. One is based on a traditional approach using hand-craft audio features and the other is a deep-learning approach using a pre-trained neural network in a transfer-learning setting. We release the \emph{Hi, KIA} dataset\footnote{\url{https://zenodo.org/record/6989810}} and the source code of the classification models\footnote{\url{https://github.com/SeungHeonDoh/hi_kia}}. We expect that they can be used for VUI-based applications in the future. 

\section{Dataset}
\emph{Hi, KIA} is a speech emotion recognition dataset for WUW. It is labeled with four emotional states including anger, happy, sad, or neutral. While this dataset was originally designed for in-vehicle VUIs, it can be also used for general-purpose  text-dependent speech emotion recognition. The entire dataset development was conducted via online platforms due to the COVID-19 pandemic situation. This section describes the step-by-step procedure.

\subsection{Scenario Selection}

The first step is selecting scenarios where the speaker utters WUW in different emotional states.
While it is possible to ask actors and actress to imagine an emotional state and speak the short words without a context, it can limit variations of nuances within the same emotional state.
To maximize the diversity and naturalness of emotional speech rendering by the voice actors and actresses, we prepared a set of user scenarios. The scenario was provided as a text script which starts with \emph{Hi, KIA} (WUW) and ends with a contextual sentence that evokes an emotional state.

For scenario selection, we worked with eight graduate students who have more than three years of experience in the area of affective computing. They were requested to come up with various situations where they use the VUI in a certain emotion. We first brought up five driver's emotions (anger, stress, happiness, fear, and sadness) based on a study by Zepf et al. \cite{zepf2020driver} and asked them to propose at least two scenarios for each emotion. As a result, we collected 53 scenarios after merging duplicated ones. We mapped them to the valence-arousal coordination as shown in Figure \ref{fig:corpus}(a).
Subsequently, we excluded the emotion category of the 4th quadrant in the emotion circumplex as there are few scenarios, and clustered the entire scenarios into angry (low valence-high arousal), happy (high valence-high arousal), sad (low valence-low arousal), and neutral (mid valence-mid arousal). 
Then, we selected three representative scenarios for each group, which are colored in red, orange, blue, and black in Figure \ref{fig:corpus}(a). 

We prepared cards for the selected scenarios to facilitate the recording process. In each card, a sentence is presented as a recording guide on top of an illustrated car interior. Additionally, a reference image was provided as the background to describe the situation of the corresponding sentence as shown in Figure \ref{fig:corpus}(b). Then, we could augment the scenario by providing an image that visualizes the situation. The visual cue is a simple yet effective method to motivate the participants (voice actors and actresses in our case) to immerse themselves into the given scenario easily \cite{alcamo2008chapter, ogilvy2004plotting}.

\subsection{Recording and Post-processing}


We recruited four voice actors and four voice actresses online who have similar recording conditions. Their average age was 31.38 years with a standard deviation of 3.90 years. The recording process was as follows: 1) We provided a recording tutorial online. In the tutorial, we provided the voice actors and actresses with an instruction to express intended emotions to be elicited. In addition, we requested them to place their mouse position 30 cm away from the microphone during the recording. After the first recording session, we provided feedback on the quality of the recorded audio to ensure that their emotions were captured correctly. Finally, the voice actors and actresses were asked to record five more utterances. As a result, we collected 576 audio files for the 12 scenarios. 
After we collected the voice recordings of complete sentences, we cropped out the WUW segment at the beginning of the recorded audio.



\begin{table}[!t]
\centering
  \caption{Label-wise number of utterance and average length}
  \label{tab:overview}
  \begin{tabular}{l|rr}
  \toprule
\textbf{Emotion Label}  & \textbf{\# utterances} & \textbf{Avg.length(s)} \\ \midrule
Angry     & 107 & 0.610        \\ 
Happy   & 129 & 0.586     \\
Sad & 133 & 0.776        \\
Neutral   & 119 & 0.588       \\ \midrule
Total     & 488 & 0.644     \\ \bottomrule
    \end{tabular}
     \vspace{3mm}
\end{table}


\subsection{Human Validation}
We conducted human validation to remove improper data from the collected recordings. The eight graduate students who participated in scenario selection conducted the evaluation. 
Given all 576 recordings presented in a random order, they classified each recording into `angry', `happy', `sad', and `neutral'. If the recording was difficult to recognize, it was classified as `unknown'. From the validation result, we removed 88 recordings that all human evaluators predicted differently from the true label. This resulted in 488 recordings as the final dataset as shown in Table \ref{tab:relatedData}. 

\begin{figure}[!t]
    \centering
    \includegraphics[width=0.9\linewidth]{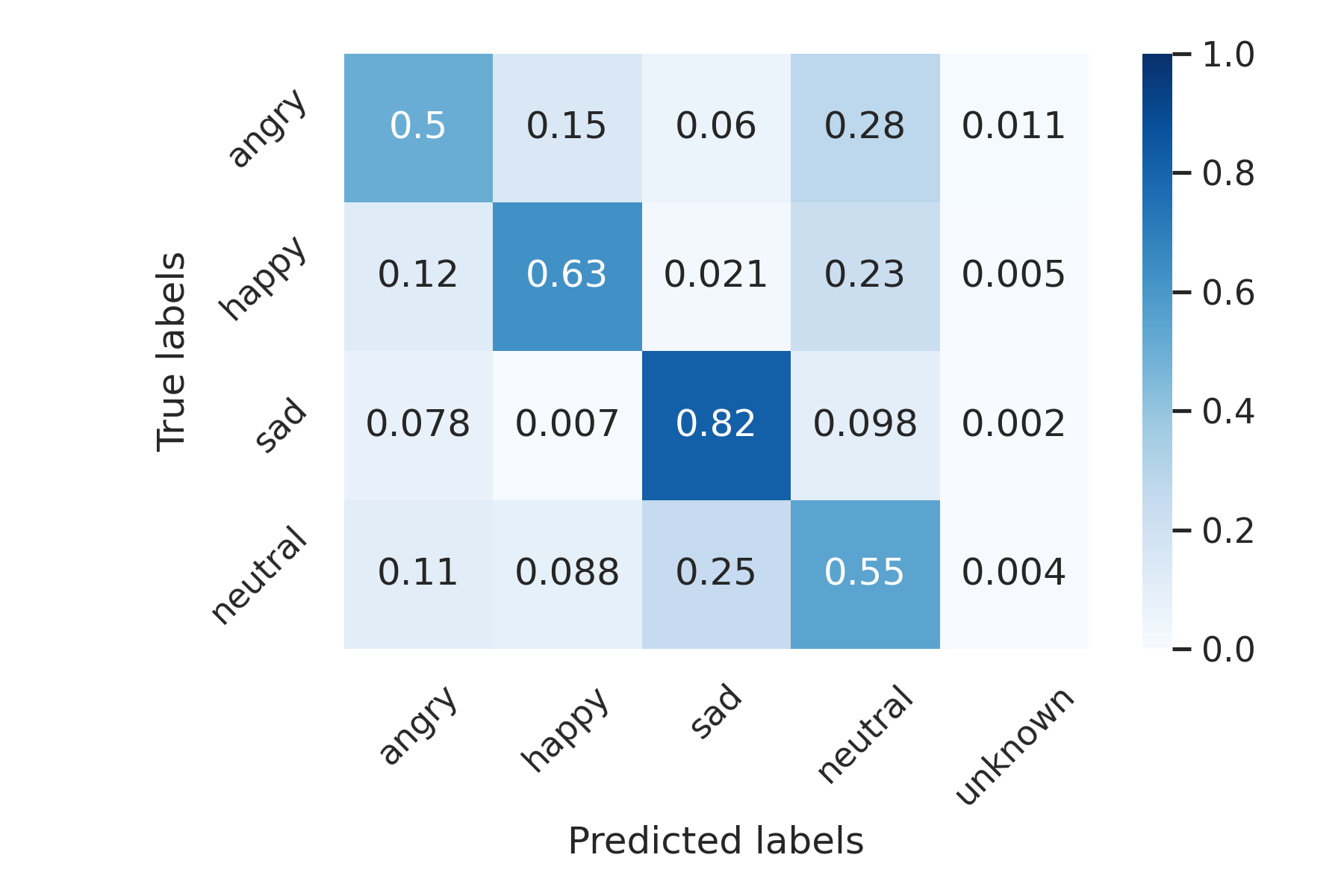}
    \caption{Confusion matrix of human validation after removing improper data.}
    \label{fig:human}
\end{figure}

We combined the human evaluators' responses and calculated the confusion matrix as shown in Figure \ref{fig:human}. We found that the evaluators were relatively good at classifying `sad' emotion. On the contrary, they had difficulty identifying `angry' and `neutral' voices; they instead evaluated `angry' as `neutral' and `neutral' as `sad'. We also noticed that they felt a high-arousal voice as emotional ground states: participants observed `angry' and `happy' as `neutral' emotions. It indicated that people have difficulty recognizing high-arousal impressions from WUW.

\vspace{3mm}

\section{Wake-up Word Emotion Recognition}

We define WUW emotion recognition as a task that predicts the speaker's emotional state from WUW utterances. 
We conducted the emotion recognition task using \emph{Hi, KIA}. 
Considering the small size of the dataset, we explored two training strategies. One is using hand-craft audio features based on domain knowledge. The other is fine-tuning a pretrained neural network model with the small dataset by leveraging the generalization capability of the model trained with a large-scale dataset. 

%

\subsection{Hand-craft Features}

\begin{figure}[!t]
     \centering
     \includegraphics[width=0.7\linewidth]{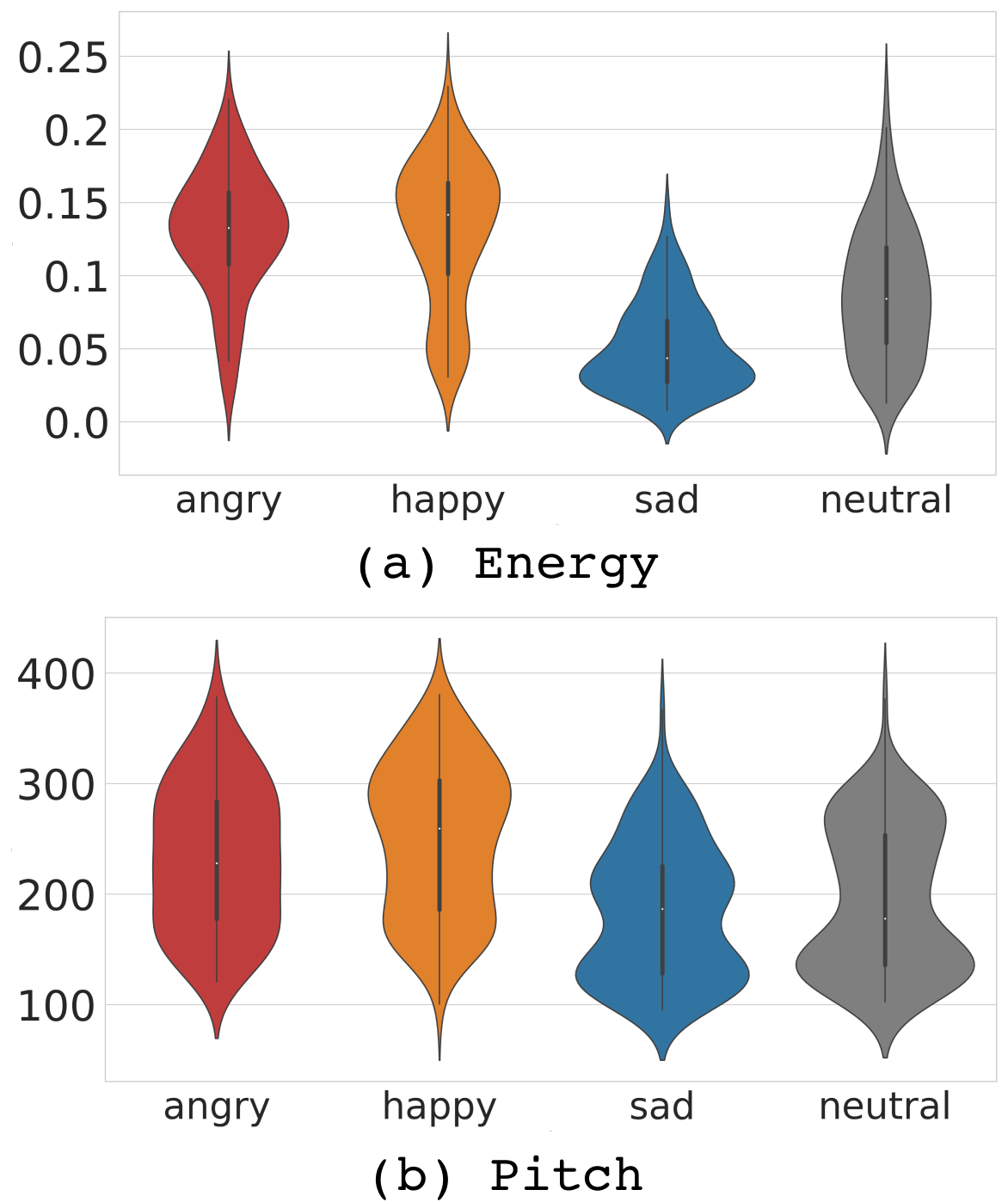}
  \caption{Utterance-level energy and pitch distribution of WUW over four emotions}
     \label{fig:hand_craft}
 \end{figure}

Speech emotion recognition is related to various acoustic properties of speech, including pitch, loudness and timbre. Traditional approaches used low-level descriptors (LLDs) or high-level statistical functions from the speech signals as input features for the classification \cite{kwon2003emotion, shen2011automatic, el2011survey}. we used the extended
\textit{Geneva Minimalistic Acoustic Parameter Set} (eGeMAPS) \cite{eyben2015geneva}, which include frequency, energy, and spectral domain features. The 88-dimensional eGeMAPS feature was z-score standardized using fixed mean and standard deviation values. Figure \ref{fig:hand_craft} shows two violin plots of energy and pitch distributions from the entire dataset. The general trend shows that the high-arousal group (`angry', `happy') is distinguished well from the low arousal group (`sad', `neutral'). For comparison with deep neural network based features, utterance-level eGeMAPS features were used as input of a logistic regression classifier. 

\subsection{Fine-Tuning with Pretrained Wav2vec 2.0}
More recent approaches use deep learning for speech emotion recognition  \cite{sarma2018emotion}. However, the lack of annotated data has limited the approaches. To address the issue, transfer learning using a large pre-trained neural network, such as Wav2vec \cite{schneider2019wav2vec}, improved the emotion recognition accuracy \cite{chen2021exploring, cai2021speech, pepino2021emotion, xia2021temporal}. With \emph{Hi, KIA}, we also conducted a similar transfer learning using Wav2vec2.0. 

\vspace{3mm}

\noindent \textbf{Pretrained Wav2vec2.0}  
Wav2vec2.0 \cite{baevski2020wav2vec} is a transformer-based model trained to extract meaningful representations from raw audio signals. Wav2vec2.0 is composed of a local encoder based on CNN, a context network based on transformer, and a quantization module. The local encoder extracts low-level representations directly from raw waveforms. Based on this representation, the context network is trained to predict future representations from past ones using a contrastive loss. The output of the context network is learned high-level representations.

\vspace{3mm}
\noindent \textbf{Fine-tuning methods}
We extracted features using Wav2vec2.0 and obtained utterance-level features through average pooling. Following the previous work \cite{xia2021temporal}, we explored different fine-tuning strategies for the modules in Wav2vec2.0. The first is measuring the emotion recognition performance in a vanila Wav2vec2.0 trained with a long-length libri-speech corpus without emotional supervision (no fine-tuning). The second is fine-tuning either the encoder network or the context network, which is responsible for a low-level or high-level representation, respectively. The last is fine-tuning the entire networks.

\subsection{Experiment Setup}

\noindent \textbf{Data split and Metrics} The \emph{Hi, KIA} dataset contains 488 clips of speech from four male and four female speakers. For speaker independence, we performed 8-fold cross-validation, where utterances from seven speakers are used for train, validation set, and utterances from the remaining one speaker are used for test set in each fold. As a result, we report both weighted accuracy (\textbf{WA}, the overall accuracy across all classes) and unweighted accuracy (\textbf{UA}, the average of the accuracy for each of the classes).  

\vspace{3mm}
\noindent \textbf{Hyper-parameters}  
We conducted experiments using the pretrained model: wav2vec2.0-base. The wav2vec-2.0-base was composed of 12 transformer blocks and 7 convolutional blocks (each has 512 channels). Our implementation was based on the Huggingface transformers repository \cite{wolf2020transformers}. We optimized the model using AdamW \cite{loshchilov2017decoupled} with parameters $\beta_{1}$=0.9, $\beta_{2}$=0.999. The learning rate used for training was $5e^{5}$, the epoch was 200. We used full audio with 16,000 Hz sampling rate, and 1 batch size. 

\begin{table}[!t]
\centering
\caption{Classification Results. `FT' stands for fine-tuning}
\label{tab:results}
\begin{tabular}{lll}
\toprule
\textbf{Model} & \textbf{WA(\%)} & \textbf{UA(\%)} \\ \midrule
Baseline (eGeMAPS) & 51.02\% & 50.63\% \\ \midrule
Wav2vec2.0 No FT & 33.61\% & 32.16\% \\
Wav2vec2.0 FT Encoder Net & 46.52\% & 44.25\% \\
Wav2vec2.0 FT Context Net & \textbf{68.64\%} & \textbf{68.51\%} \\
Wav2vec2.0 FT All & 57.99\% & 57.22\% \\ \midrule
Human Validation Performance & 63.17\% & 62.31\% \\ \bottomrule
\end{tabular}
\end{table}

\vspace{3mm}

\section{Results}
Table \ref{tab:results} shows classification performance of classification models and human validation. In the case of human validation, the final score was calculated as the average performance of 8 human evaluators. Wav2vec2.0 feature without fine-tuning does not perform better than hand-craft features. This indicates that it is difficult to extract high-level emotion features only with self-supervised learning. Fine-tuning Wav2vec2.0 significantly improves the classification accuracy. Among the three setups, fine-tuning the contextual network works best, achieving 68.64\% in WA and 68.51\% in UA. This indicates that, for a small dataset, it is more efficient to update parameters related to high-level representations rather than update all parameters. Another interesting result is that fine-tuning the contextual network outperforms human validation. This is presumably due to the subjective nature of emotion recognition. 

\begin{figure}[!t]
    \centering
    \includegraphics[width=\linewidth]{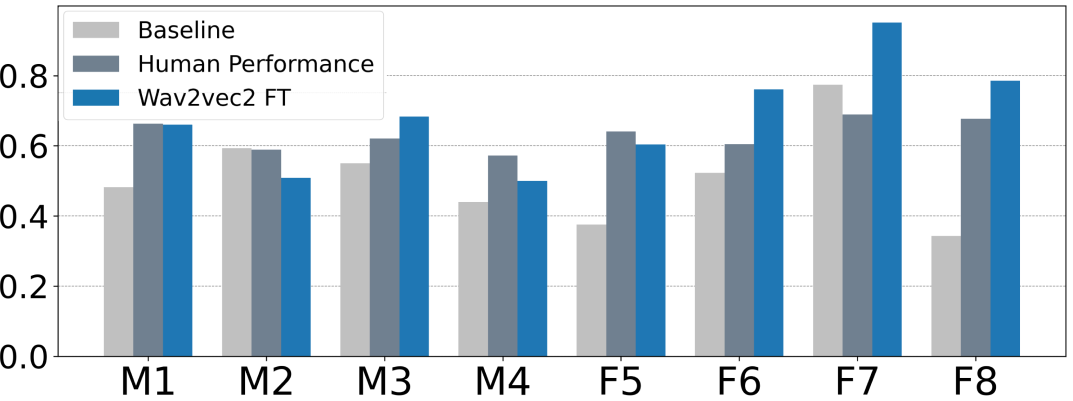}
    \caption{Fold-wise WA performance. \textsc{Wav2vec2 FT} means the result of fine-tuned context network.}
    \label{fig:foldwise}
\end{figure}

\begin{figure}[!t]
    \centering
    \includegraphics[width=\linewidth]{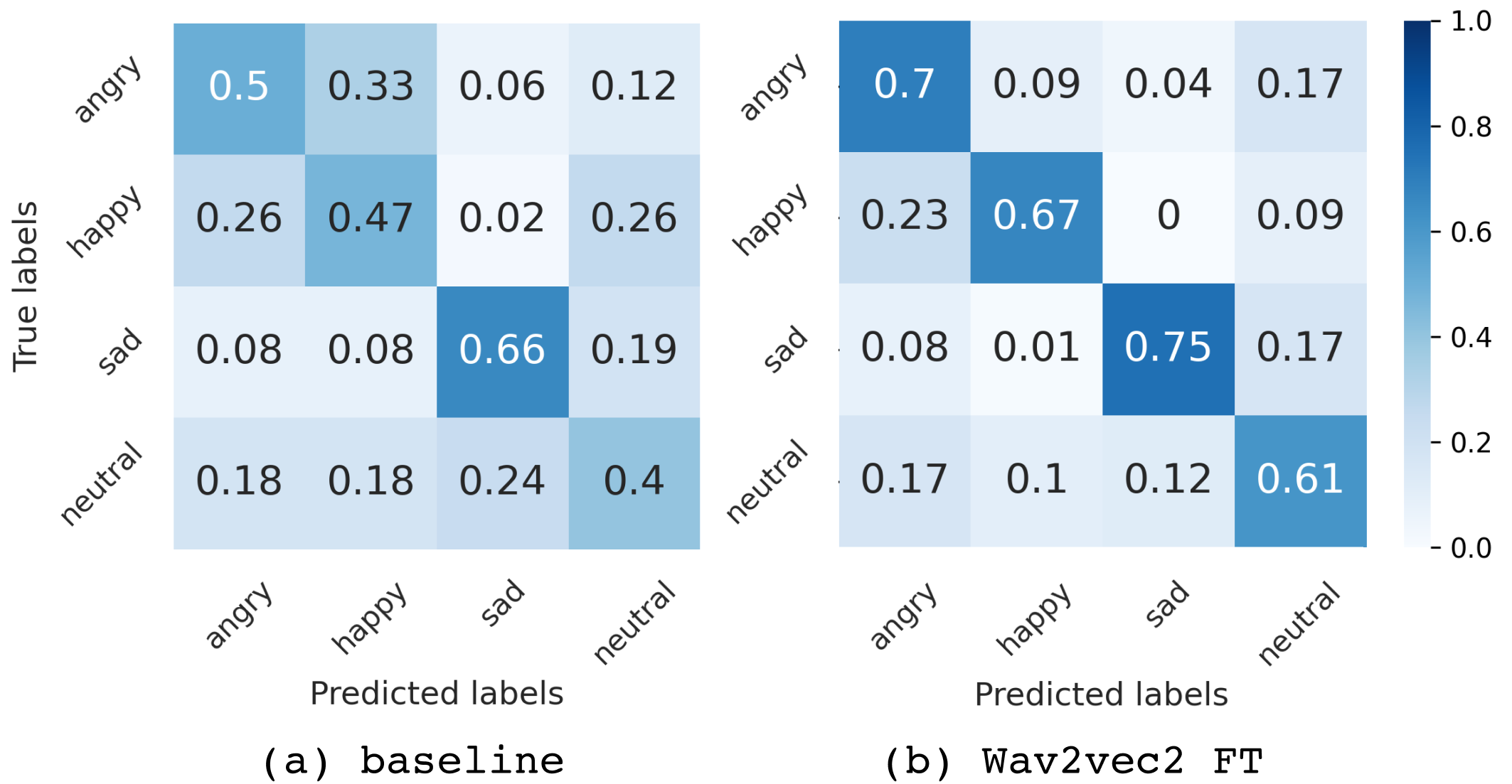}
    \caption{Confusion Matrix of \textsc{Baseline} and \textsc{Wav2vec2.FT}. WAV2VEC2.FT means the result of Fine-tune Context Network.}
    \label{fig:cm}
\end{figure}


Figure \ref{fig:foldwise} shows WA of 8 folds with 4 males and 4 females. What is noteworthy here is that the the wav2vec2.0 feature outperforms human validation performance in most female folds. Human validation performance is relatively stable in both male and female folds. Hand-craft features and `Wav2vec2.0 FT' shows a performance gap between male and female fold, especially lower performance than human validation in fold M1, M2, M4 and F5. 

Figure \ref{fig:cm} shows the confusion matrix by hand-craft feature and Wav2vec2.0 contextual network fine-tuning. Both models are good at discriminating arousal and valence differences (`happy', `sad') but hand-craft features are weak in understanding valence differences within high arousal (`angry', `happy') and sad-neutral label pairs. This problem is alleviated in Wav2vec2.0 contextual network fine-tuning. Compared with Figure \ref{fig:human} and Figure \ref{fig:cm}, Wav2vec2.0 contextual network fine-tuning outperforms human validation in discriminating high arousal emotion and neural.

\vspace{3mm}
\section{Conclusions}

This paper proposes a new public dataset, \emph{Hi, KIA}, an emotion-labeled WUW dataset. We described a carefully-designed procedure to collect short emotional utterances. After conducting human validation, we finalize the dataset composed of 488 recordings. It is a shot-length speech dataset that contains the Korean accent and utterance-level emotion annotations with four emotion classes. We have also presented baseline results for short utterance-level speech emotion recognition on this dataset, using hand-crafted features and transfer learning to overcome the limitation of a small dataset. The results show that we can achieve high accuracy in the four-way emotion recognition. As future work, we will add data on more type of emotions, such as calm or relax emotions. Then, we will conduct in-depth analysis of essential audio features to better understand WUW emotion classification. In addition, we plan to develop a speech emotion recognition model robust to speaker and gender.

\vspace{3mm}

\section*{Acknowledgement}
This research was supported by the National Research Foundation of Korea (NRF) funded by the Ministry of Education (MOE) (No. 4120200913638)
\vspace{3mm}

\bibliographystyle{IEEEtran}

\bibliography{mybib}


\end{document}